\documentclass[preprint,12pt]{elsarticle}

\usepackage[usenames]{color}

\usepackage{amssymb}

\newtheorem {theo} {\bf Theorem} [section]
\newtheorem {prop} [theo] {\bf Proposition}
\newtheorem {cory} [theo] {\bf Corollary}

\newtheorem {defn} [theo] {\bf Definition}
\newtheorem {conj} [theo] {\bf Conjecture}

\newcommand{\QED}{\hfill \CaixaPreta \vspace{6mm}}
\def\CaixaPreta{\vrule Depth0pt height6pt width6pt}

\newcommand{\ot}{\otimes}

\usepackage{graphicx}

\begin{document}

\begin{frontmatter}



\title{Impulse Stability of Large Flocks: an Example}

\author[jjpv]{ J. J. P. Veerman}
\ead{veerman@pdx.edu}

\author[fmt]{F. M. Tangerman\corref{cor1}}
\ead{fmtangerman@gmail.com}

\cortext[cor1]{Corresponding Author}

\address[jjpv]{Dept. of Math. \& Stat., Portland State
University, Portland, OR 97201, USA.}

\address[fmt]{Dept. of Math, Stony Brook University,
Stony Brook, NY 11794-3651, USA.
Tel: 1-631-632-8250, Fax: 1-631-632-7631.}



\begin{abstract}
Consider a string of $N+1$ damped oscillators moving in $\mathbb{R}$ of which the motion of the first (called the ``leader") is independent of the others. Each of the followers `observes' the relative velocity and position of only its nearest neighbors. Inasmuch as these are different from 0, this information is then used to determine its own acceleration. Fix all parameters \emph{except} the number $N$ in such a way that the system is asymptotically stable. Now as $N$ tends tends we consider the following problem. At $t=0$ the leader gets kicked and starts moving with unit velocity
away from the flock. Due to asymptotic stability the followers will eventually fall in behind the leader and travel each
at its own predetermined distance from the leader. In this note we conjecture that before equilibrium ensues,
the perturbations to the orbit of the last oscillator grow \emph{exponentially} in $N$ \emph{except} when there is a symmetry in the interactions and the growth is then \emph{linear} in $N$. There are two cases. We prove the conjecture
in one case, and give a strong heuristic argument in the other.
\end{abstract}


\end{frontmatter}


\vskip 1.in
\section{Introduction} \label{chap:intro}

\setcounter{figure}{0} \setcounter{equation}{0}

In this note we study how a long, but finite, string of asymmetrically coupled damped oscillators reacts as one of its members (the \emph{leader}) changes its velocity suddenly. The aim is to study how the stability of large flocks depends on the kind of interaction between the individual agents.

\begin{figure}[ptbh]
\centering
\includegraphics[width=5.0in]{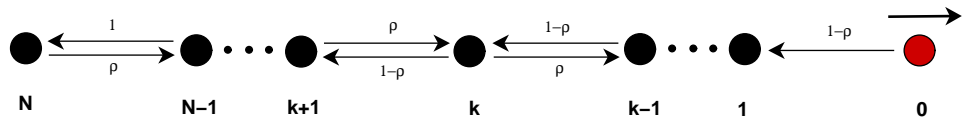}
\caption{\emph{ The communication graph of the system. Each agent is linearly coupled to its nearest neighbor. The arrows give the direction of the information flow. The interactions for the boundary agents are different from those in the interior.
At $t=0$ the agent labeled 0, the leader, undergoes a forced motion: a kick in the direction of the arrow above it. It receives no feedback from the flock.  }}
\label{fig:comm-graph}
\end{figure}

Assume then that we have a string of oscillators as depicted in Figure \ref{fig:comm-graph}, where in principle each agent observes the relative distance and velocity of its immediate neighbors and uses
those to compute its own acceleration. The types of interaction are indexed by a parameter $\rho\in[0,1]$: each agent multiplies the information coming from the neighbor `in front' by weight $1-\rho$ and from the neighbor `behind' by the weight $\rho$. Naturally the interactions of the first and last agents are a little different from those with two neighbors. The leader itself is assumed not at all influenced by the rest of of the flock (an `independent leader'). At $t=0$ the agent labeled 0, the leader, undergoes a forced motion: namely its velocity goes instantaneously from 0 to 1 (it receives a `kick').
The response of the last agent is called the 'impulse response' function. We study this response function holding all parameters fixed,
except the number $N$ of agents.

The interplay between graph theory and coupled linear ODE's is described in detail in \cite{flocks2}. In particular,
the collection of agents connected by the arrows that give the information flow gives a directed graph, known as the \emph{communication graph}, see Figure \ref{fig:comm-graph}.

In \cite{flocks4, flocks5} we studied what happens if the weights are equal, that is: $\rho=\frac12$. In that case the perturbation propagates from the leader throughout the flock and is roughly multiplied by $N$, the size of the flock, when it arrives at the trailing agent (labeled $N$). The perturbation then decays as the system is assumed to be asymptotically stable. In \cite{flocks6} and this paper we take up the study of this system when $\rho\neq\frac12$. In the former we concluded that the \emph{frequency response} function grows exponentially for all $\rho\in [0,1)$. Here we we show that if $\rho\in(\frac12,1)$, then the impulse response function also shows exponential growth in $N$. When $\rho\in (0,\frac12)$ we conjecture that this is also true. (When $\rho=\frac12$, both grow only linearly with $N$).

We study systems of the form (for details see \cite{flocks6}):
\begin{equation}
\dot z = M z + \Gamma_0(t)\quad ,
\label{eq:indepleader}
\end{equation}
where $z=(z_1,\,\dot{z}_1,...,z_N,\,\dot{z}_N)$ is the vector of position and velocity of the agents. The leading car is not encoded since its
orbit $z_0(t)$ is a priori given.  The matrix $M$ is defined in terms of the Kronecker product ($\otimes$)
 \begin{equation}
M\equiv I \ot A  + P \ot K \quad  .
 \end{equation}
Here $I$ and $P$ are $N$-dimensional square matrices, where $I$ is the identity and $P$ is given by
\begin{equation}
P= I-Q_\rho \quad where\quad Q_\rho=\left(\begin{array}{ccccc}
0 & \rho & & & \\
1-\rho & 0 & \rho & & \\
 & \ddots & \ddots &  \ddots & \\
 & & 1-\rho & 0 & \rho \\
 & & & 1 & 0
\end{array}\right),
\label{eq:laplacian}
\end{equation}
The $2\times2$ matrices $A$ and $K$ are given by:
\begin{equation}
A= \left(\begin{array}{cc}
0 & 1 \\
0 & 0
\end{array}\right) \quad and \quad
K= \left(\begin{array}{cc}
0 & 0 \\
f & g
\end{array}\right) \quad .
\label{eq:A-and-K}
\end{equation}
Finally:
\begin{equation}
\Gamma_0(t) = \left(\begin{array}{c}
0  \\
(1-\rho)\left(fz_0(t)+g\dot z_0(t)\right) \\
0\\
\vdots
\end{array}\right) \quad  .
\label{eq:Gamma_0}
\end{equation}

We consider the problem where the flock is at equilibrium for time $t < 0$, that is: the agents are at rest and properly spaced. At time $0$ we provide an impulse to the leader: $\ddot{z_0}(t)=\delta_0(t)$, the leader then advances at constant velocity equal to $1$ for time $t>0$.
This problem is ultimately motivated by what is called the \emph{canonical traffic problem} in \cite{flocks4, flocks5}, in which one imagines a long row of cars waiting for the traffic light to turn green. When that happens, the first car, the leader, quickly accelerates to the desired speed and the others aim to follow it. The problem now is to deduce the motion of the other agents $z_k(t)$, $k=1,...,N$ and $t>0$.

The control parameters $f,g < 0$, are assumed negative, implying asymptotic stability (\cite{flocks6}), so the flock ultimately follows the leader in equilibrium formation. Here we isolate the following problem: What is the transient behavior of $z_k(t)$? The orbit of the agent farthest from the leader, $z_N(t)$, is plausibly the one that suffers the worst effects (and this is amply born out by numerous numerical experiments). To simplify the discussion we concentrate on its orbit.

This problem constitutes part of a larger research project that aims to analytically understand the dynamics of a large number of agents trying to move coherently in a changing environment that causes a few elements, the leaders, to react to it. An interesting class of systems is given by the requirements that most agents ``receive information" from the same number of nearby agents (homogeneity), and that this information only consists of their neighbors' position and velocity. This pattern only changes for agents on the boundary of the flock (where it has fewer neighbors). In our case we furthermore insist on the interaction being \emph{linear} and involving only positions and velocities of the neighbors. The aim is to give a qualitative analysis of the transients of these systems (``Newtonian Networks" is a concept we propose hereby) as the number of agents is very large. The analysis below reflects some of the complications of this endeavor by examining a (apparently simple) paradigm of this idea.

\vskip .2in
\noindent{\bf Notational Conventions:} To avoid confusion, we list two important conventions here. The first is that we assume that both $f$ and $g$ are negative reals to insure asymptotic stability (Theorem \ref{theo:stable}). The second is that we define the symbol $\sqrt z$ as the root with angle in the interval $[0,\pi)$ (branch cut along the positive real axis).

\section{Preliminary Results} \label{chap:Zeros}

In this section we first give the eigenvalues of the matrix $M$ of of Equation (\ref{eq:indepleader}). We then give the expression for the frequency response function for the trailing agent and discuss its singularities. The following constant will frequently simplify formulae:
\begin{displaymath}
\kappa \equiv \frac{1-\rho}{\rho} \quad or \quad \rho = \frac{1}{1+\kappa} \quad .
\end{displaymath}

In the statement of the next result and that of Proposition \ref{roots} we use the following equation, where $\rho \in(0,1)$ and $\phi$ are real variables:
\begin{equation}\label{cot2}
(2\rho - 1)\cot\phi=\cot N\phi\; .
\end{equation}
Recall that the matrix $P$ is defined in Equation (\ref{eq:laplacian}).

\vskip.2in\begin{prop} (\cite{tridiagonal})
For any $\rho\in(0,1)$, the matrix $P$ has $N$ distinct eigenvalues $\{\lambda_\ell\}_{\ell=0}^{N-1}$: \\
\emph{\bf i) If $\rho\in (0,\frac{1}{2}]$:} for $\ell\in\{0,\ldots, N-1\}$, $\lambda_\ell=1-2\sqrt{\rho(1-\rho)}\,\cos \phi_\ell$, where $\phi_\ell \in \left( \frac{\ell\pi}N, \frac{(\ell+1)\pi}N\right)$ solves (\ref{cot2}).\\
\emph{\bf ii) If $\rho\in (\frac12,\frac{N+1}{2N}]$:} Identical to i).\\
\emph{\bf iii) If $\rho\in (\frac{N+1}{2N},1)$:} for $\ell\in\{1,\ldots, N-2\}$, $\lambda_\ell=1-2\sqrt{\rho(1-\rho)}\,\cos \phi_\ell$, where $\phi_\ell \in \left( \frac{\ell\pi}N, \frac{(\ell+1)\pi}N\right)$ solves (\ref{cot2}); $\lambda_0=\frac{(2\rho-1)^2}{2\rho^2}\left(\frac{1-\rho}{\rho} \right)^{N-1} + \mathcal{O}\left(\left(\frac{1-\rho}{\rho} \right)^{2N-2}\right)$ and $\lambda_{N-1}=2-\lambda_0$.
\label{prop:evalsP}
\end{prop}

\noindent One can show (see \cite{flocks2,flocks4, flocks5}) that the eigenvalues of $M$ defined in Equation (\ref{eq:indepleader}) are given by the solutions $\nu_{\ell\pm}$ of
 \begin{equation}
 \nu^2-\lambda_\ell g \nu-\lambda_\ell f= 0  \quad,
  \label{eq:evals2}
 \end{equation}
where $\lambda_\ell$ runs through the spectrum of $P$. So:

\vskip.2in
\begin{theo} The eigenvalues of $M$ are
\begin{displaymath}
\nu_{\ell\pm} = \frac{1}{2}\left(\lambda_\ell g \pm \sqrt{(\lambda_\ell g)^2 + 4 \lambda_\ell f}\right)= \frac{\lambda_\ell g}{2}\left(1\pm \sqrt{1+\frac{4f}{\lambda_\ell g^2}}\right) \quad ,
 \end{displaymath}
where $\lambda_\ell$ runs through the spectrum of $P$. Because the $\lambda_\ell$ are contained in the interval $[0,2]$ (see Proposition \ref{prop:evalsP}), the system is stabilized (or globally stable) if and only if both $f$ and $g$ are strictly smaller than zero.
\label{theo:stable}
\end{theo}

\vskip.2in
\begin{cory} The eigenvalues $\nu_{\pm\ell}$ of $M$ in the complex $\nu$ plane  either lie on the circle $|\nu+\frac{f}{g}|^2=\frac{f^2}{g^2}$, namely whenever $\frac{4|f|}{\lambda_\ell g^2}>1$, or else are real numbers less than or equal to $-\frac{|f|}{|g|}$.
\label{cory:location-evals}
\end{cory}

Now we turn to the frequency response function of the trailing car when $\rho\neq\frac12$.

\vskip.2in
\begin{cory} (\cite{flocks6}) For $\rho\in (0,1)\backslash \{\frac{1}{2}\}$ the frequency response function of the last agent is given by
\begin{displaymath}
\begin{array}{c}
a_{N}(\nu)=  \frac{1+\kappa}{\kappa} \;\kappa^N \; \frac{\mu_+ - \mu_-}
{\left(\mu_+-\mu_+^{-1}\right)\mu_+^{N}- \left(\mu_--\mu_-^{-1}\right)\mu_-^{N}} \quad ,\\
where \quad  \mu_\pm=\mu_\pm(\nu) \equiv \frac{1}{2\rho}\left(\gamma\pm \sqrt{\gamma^2-4\rho(1-\rho)}\right) \quad and \quad
\gamma = \gamma(\nu)\equiv  \frac{f+ g \nu -\nu^2}{f+ g\nu}
\quad .
\end{array}
\label{cor:trailing}
\end{displaymath}
As functions of $\nu$, the $a_k(\nu)$ in fact are proper rational functions.
\end{cory}

\vskip.2in
In what follows the location of the roots of the denominator of $a_N$ is important. Recall that $\mu_+\mu_-=\kappa$ and define the function
$f:\mathbb{C} \rightarrow \mathbb{C}$
equal to the denominator:
\begin{displaymath}
f(\mu)\equiv \mu^{N+1}-\mu^{N-1}-\left(\left(\frac{\kappa}{\mu}\right)^{N+1} - \left(\frac{\kappa}{\mu}\right)^{N-1}\right) \quad .
\end{displaymath}

We note that we have two different representations of the set
$\nu_{\pm \ell}$, one in terms of eigenvalues $\lambda_{\ell}$ and the other in terms of roots $\mu_{\pm \ell}$. The locations of the zeros $\mu_{\pm \ell}$ are described in the following result

\vskip.2in
\begin{prop} (\cite{tridiagonal}) \label{roots}
For any positive real number $\kappa$, the function $f$ has
$2N+2$ roots. Two of these are the fixed points of the involution $h:\mu\rightarrow \mu=\frac{\kappa}{\mu}$ and are given by
$\pm \sqrt{\kappa}$. The remaining $2N$ roots, $\{\mu_{\ell\pm}\}_{\ell=0}^{N-1}$, have period $2$ under the involution $h$ and are given as follows: \\
\emph{\bf i) If $\kappa\geq 1$:} $N$ roots are given by $\mu_{\ell+}=\sqrt\kappa \, e^{i\phi_\ell}$, where $\phi_\ell \in \left( \frac{\ell\pi}N, \frac{(\ell+1)\pi}N\right)$, for $\ell\in\{0,\ldots, N-1\}$, solves (\ref{cot2}); the remaining roots are the images under $h$ of these or: $\sqrt\kappa \, e^{-i\phi_\ell}$, respectively.\\
\emph{\bf ii) If $\kappa\in [\frac{N-1}{N+1},1)$:} Identical to i).\\
\emph{\bf iii) If $\kappa\in (0,\frac{N-1}{N+1})$:} $N-2$ roots are given by $\mu_{\ell+}=\sqrt\kappa \, e^{i\phi_\ell}$, where $\phi_\ell \in \left( \frac{\ell\pi}N, \frac{(\ell+1)\pi}N\right)$, for $\ell\in\{1,\ldots,N-2\}$, solves (\ref{cot2}); $N-2$ are images of these under $h$; the remaining roots are $\mu_{0+}\in(\sqrt{\kappa},1)$ and its images under $h$ and multiplication by -1. We have $\mu_{0+}= 1 -\frac{1}{2}(1-\kappa^2)\kappa^{N-1} + \mathcal{O}(\kappa^{2N-2})$.\\
\end{prop}

\vskip 0.5in
\section{Laplace transform and Residues} \label{chap:time}

In Equation (\ref{eq:indepleader}) we set $\ddot z_0(t)=\delta (t)$. Set the initial conditions as follows: for all $k\geq 1$: $\dot z_k(0)=z_k(0)=0$. Then $\ddot z_k(t)$ equals the Green's function of this problem (see \cite{flocks5}):
\begin{equation}
\ddot z_k(t) \equiv \frac{1}{2\pi i}\,\int_{r-i\infty}^{r+i\infty}\;a_k(\nu)e^{\nu t}\,d\nu \quad .
\label{eq:invlaplace}
\end{equation}
The actual impulse response functions $z_k(t)$ can be obtained from this by twice integrating with the usual initial conditions ($\dot z_k(0)=z_k(0)=0$). Here we calculate the impulse response of the last car (labeled by $N$). The strategy is to perform a residue expansion (or partial fraction expansion) of $a_N(\nu)$ (given in Corollary \ref{cor:trailing}).

Considering Corollary \ref{cor:trailing} we write $a_N(\nu)=\frac{p(\nu)}{q(\nu)}$ as a quotient of polynomials with degree($p$) at least that of degree($q$). The zeros of $q$ are the eigenvalues of $M$. Thus
according to Theorem \ref{theo:stable} the denominator in $a_N(\nu)$ has only simple roots located at $\nu_i$, except when for some $\ell$: $-4f=\lambda_\ell g^2$. Avoiding that case for simplicity, we have:
\begin{displaymath}
a_N(\nu)=\sum_i \frac{{\rm Res}(a_N,\nu_i)}{\nu-\nu_i} \quad where \quad  {\rm Res_N}(a,\nu_i)= \frac{p(\nu_i)}{q'(\nu_i)} \quad .
\end{displaymath}
With this proviso, we will calculate all the residues ${\rm Res}(a_N(\nu),\nu_{\pm \ell})$, $\nu_{-\ell}=\overline{\nu_{\ell}}$ and
\begin{equation}
a_N(\nu) = \sum_{\ell=0}^{N-1}
\left(\frac{{\rm Res}(a_N(\nu),\nu_{\ell-})}{\nu-\nu_{\ell-}} + \frac{{\rm Res}(a_N(\nu),\nu_{\ell+})}{\nu-\nu_{\ell+}} \right) \quad.
\label{eq:residue}
\end{equation}
The indexing has been chosen so that the pair $\nu_{\pm \ell}$ correspondings to a pair $\mu_{\pm \ell}$ of zeroes of $f$.

This representation allows us to 'compute' the motion of the N-th agent via the inverse Laplace transform
\begin{equation}
z_N(t)= \sum_{\ell=0}^{N-1}
\left(\frac{{\rm Res}(a_N(\nu),\nu_{\ell-})}{\nu_{\ell-}^2}\;e^{\nu_{\ell-}t}+ \frac{{\rm Res}(a_N(\nu),\nu_{\ell+})}{\nu_{\ell+}^2} \;e^{\nu_{\ell+}t}\right) +C_N +D_N t\quad.
\label{eq:impulse-resp.}
\end{equation}
\vspace{.2in}
\noindent
The constants of integration $C_N$ and $D_N$ have to guarantee that $z_N(0)=0$, $\dot{z}_N(0)=0$.

\vskip .2in
\begin{theo}
If the poles are simple then
\begin{displaymath}
 {\rm Res}(a_N(\nu),\nu_{\ell\pm})= -\; \frac{(f+g\nu_{\ell\pm})^2}{\nu_{\ell\pm}\,(2f+g\nu_{\ell\pm})}\; \; \frac{\kappa^{N-1} \mu_{+\ell}^{N-3}(\mu_{+\ell}^2-\kappa)^2} {2N\mu_{+\ell}^{2N-2}(\mu_{+\ell}^2-1)+ 2\mu_{+\ell}^{2N}+2\kappa^{N-1}}\quad .
\end{displaymath}
\label{theo:residue}
\end{theo}

\noindent {\bf Proof:} If in the above Proposition we replace $\mu_-$ by $\kappa/\mu_+$, then the expression for $a_N$ in Corollary \ref{cor:trailing} is a rational function of $\mu_+$ alone:
\begin{displaymath}
a_N = \frac{1+\kappa}{\kappa}\; \frac{\kappa^N \mu_+^N(\mu_+^2-\kappa)}{(\mu_+^2-1)\mu_+ ^{2N}+(\mu_+^2-\kappa^2)\kappa^{N-1}}
\equiv \frac{1+\kappa}{\kappa}\; \frac{p_N(\mu_+)}{q_N(\mu_+)} \quad .
\end{displaymath}
(The polynomials $p_N$ and $q_N$ still have a factor $(\mu_+^2-\kappa)$ in common, which is kept to simplify the calculation.) Recall that $\mu_+$ is a function of $\gamma$ by (choose the ``+"root):
\begin{displaymath}
(1-\rho)-\gamma \mu + \rho \mu^2 = 0 \quad ,
\end{displaymath}
with $\gamma$ determined by $\nu_{\ell\pm}$ through
\begin{displaymath}
\gamma=1-\frac{\nu^2}{f+g\nu} \quad .
\end{displaymath}
The pole expansion of $a_N$ is performed as in \cite{flocks5}.  When
the poles of $a_N$ are simple we obtain that
\begin{displaymath}
{\rm Res}(a_N(\nu),\nu_{\ell\pm})=\frac{1}{\mu_+'(\nu_{\ell\pm})} {\rm Res}(a_N(\mu),\mu_{\ell\pm})
\end{displaymath}
Using the above relations, one obtains:
\begin{displaymath}
\mu_+'(\nu_{\ell\pm})= -(1+\kappa)\,\frac{\mu_+(\nu_{\ell\pm})^2}{\mu_+(\nu_{\ell\pm})^2-\kappa} \; \frac{\nu_{\ell\pm}(2f+g\nu_{\ell\pm})}{(f+g\nu_{\ell\pm})^2} \quad .
\end{displaymath}
Using this and replacing the residue of $a_N(\mu_+)$ by $\frac{1+\kappa}{\kappa}\,\frac{p_N(\mu_{+\ell})}{q_N'(\mu_{+\ell})}$:
\begin{eqnarray*}
{\rm Res}(a_N(\nu),\nu_{\ell\pm})&=&  -(1+\kappa)^{-1} \left(\frac{\mu_{+\ell}^2-\kappa}{\mu_{+\ell}^2}\right) \;
\frac{(f+g\nu_{\ell\pm})^2}{\nu_{\ell\pm}\,(2f+g\nu_{\ell\pm})}.\\
&&\frac{(1+\kappa)\kappa^{N-1} \mu_{+\ell}^{N}(\mu_{+\ell}^2-\kappa)} {2N\mu_{+\ell}^{2N-1}(\mu_{+\ell}^2-1)+ 2\mu_{+\ell}(\mu_{+\ell}^{2N}+\kappa^{N-1})}\quad ,
\end{eqnarray*}
which after some simplification gives the desired result.
\QED

\vskip 0.5in
\section{The Dominant Poles} \label{chap:condition1}

We show that there are three cases:
\begin{enumerate}
\item When $\rho> 1/2$ (or $\kappa<1$), the weight is more on the agent following. In this case two poles dominate the frequency response $a_N(\nu)$ and we can estimate the impulse response by the inverse Laplace transform.
\item When $\rho<1/2$ (or $\kappa>1$), the weight favors the agent in front. In this case no poles appear to be negligible, and the inverse transform is problematic.
\item When $\rho=1/2$ (or $\kappa=1$), equal weight is on the front and back neighbor. In this case on the order $\sqrt N$ poles dominate. The inverse transform can be done and this case is described in \cite{flocks5}.
\end{enumerate}

Theorem \ref{theo:stable} implies that the sign of $1+\frac{4f}{\lambda_\ell}$ determines whether the eigenvalues of the system are real or complex. Proposition \ref{prop:evalsP} tells us that when $\kappa<1$ the eigenvalue $\lambda_0$ is exponentially small. Given our assumptions (all parameters fixed, except $N\rightarrow\infty)$ we may thus assume that $\nu_{0\pm}$ are complex when $\kappa<1$. When $\kappa>1$ it is certainly possible that \emph{all} eigenvalues are real. To simplify the analysis we will assume from now on that all eigenvalues are pairs of complex conjugates (or that $2\frac{|f|}{|g|^2}>1$).

We now present the magnitudes of the residues as well as the relative location of the eigenvalues $\nu_{\pm \ell}$ to the real axis in terms of a table. First, define the following factors:
\begin{displaymath}
\begin{array}{ccc}
I&\equiv & -\;\frac{(f+g\nu_{\ell\pm})^2}{\nu_{\ell\pm}\, (2f+g\nu_{\ell\pm})}\\[.4cm]
II&\equiv& \; \; \frac{\kappa^{N-1} \mu_{+\ell}^{N-3}(\mu_{+\ell}^2-\kappa)^2} {2N\mu_{+\ell}^{2N-2}(\mu_{+\ell}^2-1)+ 2\mu_{+\ell}^{2N}+2\kappa^{N-1}}\\[.4cm]
III&\equiv&  \frac{1}{i\Im(\nu_{\ell\pm}) -\nu_{\ell\pm}}
\end{array}
\end{displaymath}

The factors $I$ and $II$ multiply to produce the residue, while factor $III$ describes the relative inverse location of the poles to the imaginary axis: $III$ is large when the poles are near the imaginary axis.
The following tables hold:

\begin{tiny}
\[
\begin{tabular}{|c||c|c|c|c|c|c|}
\hline
$\kappa<1$ &&&&&&\\
\hline
$\# $ & $\lambda$ & $\nu$ & $\mu$ & I & II & III\\
\hline \hline
0+ & $\frac12 (1-\kappa)^2 \kappa^{N-1}$ & $\frac12 \lambda_0 g + i\sqrt{\lambda_0|f|}$ & $1-\frac12 (1-\kappa^2)\kappa^{N-1}$ & $\frac{-i\sqrt{|f|}}{2\sqrt{\lambda_{0}}}$ & $\lambda_0$ & $\frac{2}{\lambda_0 |g|}$\\
\hline
0- & $\frac12 (1-\kappa)^2 \kappa^{N-1}$ & $\frac12 \lambda_0 g - i\sqrt{\lambda_0|f|}$ & $\kappa(1+\frac12 (1-\kappa^2)\kappa^{N-1})$ & $\frac{i\sqrt{|f|}}{2\sqrt{\lambda_{0}}}$ & $\lambda_0$ & $\frac{2}{\lambda_0 |g|}$\\
\hline
$\ell+$ & $1-2\sqrt{\rho(1-\rho)}\cos\phi_\ell$  & ${\cal O}(1)$ & ${\cal O}(1)$ & ${\cal O}(1)$ & ${\cal O}(\kappa^{N/2})$ & ${\cal O}(1)$\\
\hline
$\ell-$ & $1-2\sqrt{\rho(1-\rho)}\cos\phi_\ell$   & ${\cal O}(1)$ & ${\cal O}(1)$ & ${\cal O}(1)$ & ${\cal O}(\kappa^{N/2})$ & ${\cal O}(1)$\\
\hline
$(N-1)+$ & $2-\lambda_0$ & ${\cal O}(1)$ & ${\cal O}(1)$ & ${\cal O}(1)$ & ${\cal O}(\kappa^N)$ & ${\cal O}(1)$\\
\hline
$(N-1)-$ & $2-\lambda_0$ & ${\cal O}(1)$ & ${\cal O}(1)$ & ${\cal O}(1)$ & ${\cal O}(\kappa^N)$ & ${\cal O}(1)$\\
\hline
\end{tabular}
\]
\end{tiny}
\begin{tiny}
\[
\begin{tabular}{|c||c|c|c|c|c|c|}
\hline
$\kappa>1$ &&&&&&\\
\hline
$\# $ & $\lambda$ & $\nu$ & $\mu$ & I & II & III\\
\hline \hline
$\ell\pm$ & $1-2\sqrt{\rho(1-\rho)}\cos\phi_\ell$  & $\asymp 1$ & $\sqrt{\kappa}\;e^{\pm i\phi_\ell}$ & $\asymp 1$ & $\frac{2\kappa^{\frac{N+1}{2}}\; e^{-iN\phi_\ell}\;\sin^2 \phi_\ell}{N(\kappa\;e^{i\phi_\ell}-e^{-i\phi_\ell})}\left(1+{\cal O}(1/N)\right)$& $\asymp 1$\\[.3cm]
\hline
\end{tabular}
\]
\end{tiny}
\vskip .1in
(The symbol $I\asymp 1$ means that the absolute value of the expression ``I" is ${\cal O}(1)$ but not ${ o}(1)$).

\vskip .1in
The accuracy in the table is as follows. Where the entry equals $ax$ the accuracy is ${\cal O}(x^2)$, where the term is $ax+bx^2$ (such as in both the $\nu$-column and the I-column) the accuracy is ${\cal O}(x^3)$. Otherwise exceptions are mentioned. For example, one of these, the entry for
$\nu_{\ell\pm}$, follows because the absolute value of the imaginary part of $\nu_{\ell\pm}$ is ${\cal O}(1)$. Note that the $\lambda$'s necessarily have the same ``+" and ``-" entries.

We first discuss the situation when $\kappa<1$. For $\nu$ close enough to the pole at $\nu_{0+}$, $a_N(\nu)$ can be estimated by evaluating the behavior of $a_N$ near that pole (see Equation (\ref{eq:residue})). Substitute $\nu = i\sqrt{\lambda_0|f|}$ and (see Equation (\ref{eq:residue})) multiply the factors I, II, and III, of the first line in the table to obtain the following result (see also Theorem 4.6 of \cite{flocks6}):

\vskip .2in
\begin{cory} When $\kappa<1$,
$a_N(i\sqrt{\lambda_0|f|})= -i \frac{\sqrt{2|f|}}{|g|}\;\frac{\kappa^{(1-N)/2}} {(1-\kappa)}\;+{\cal O}(1)$.
\label{cor:A_N}
\end{cory}

This result should be compared with Theorem 4.6 of \cite{flocks6} (especially its proof).

From Equation (\ref{eq:impulse-resp.}) and the table above, we see that there are only two poles that yield an exponential contribution to the impulse response function $z_N(t)$. From the table one can calculate this contribution by multiplying I and II, and dividing the result by the square of $\nu$ (see Equation (\ref{eq:impulse-resp.})).

\vskip .2in
\begin{cory} Using only the two principal poles (when $\kappa<1$) as an approximation, the impulse response for the trailing car is given by:
\begin{displaymath}
z_N(t) = t-\frac{1}{\sqrt{|f|}\;\sqrt{\lambda_0}} \; e^{\lambda_0 gt/2}\;\sin(\sqrt{\lambda_0 |f|}\;t) \quad
where \quad \lambda_0 = \frac12 (1-\kappa^2) \kappa^{N-1} \quad .
\end{displaymath}
(The two leading poles determine the dynamics of the last agent.)
\label{cor:orbit}
\end{cory}

The dynamics in this case is virtually entirely determined by the leading eigenvalue $\lambda_0$ of the (reduced) Laplacian.
Note that $\lambda_0$ is exponentially small, yet positive, in N. Also notice that the constant $g$ is negative. The motion $z_N(t)$ for a substantial time interval (as long as $\sqrt{\lambda_0 |f|}\;t$ is small) is roughly equal to $z_N(t)=t(1-e^{\lambda_0 gt/2})$ and therefore remains small, that is: for an amount of time $O(\kappa^{\frac{N}{2}})$ the agent appears to not move.

Now we turn to the case where $\kappa>1$. From the tables we conclude:

\vskip .2in
\begin{cory}
 When $\kappa>1$ there are ${\cal O}(N)$ poles that play a role in the dynamics of the orbit of the trailing car. (All have exponentially large residues while other factors are ${\cal O}(1)$.)
\label{cor:many-poles}
\end{cory}

The fact that there are no dominant poles in this situation effectively prevents us from giving an approximation of the impulse response function.

\section{Impulse Stability} \label{chap:impulse}

In \cite{flocks6} we suggest the following notion of impulse stability.
\vskip .2in
\begin{defn} Consider Equation (\ref{eq:indepleader}) with forcing determined by $\ddot z_0(t)\equiv \delta(t)$ and subject to the initial conditions $z_k(0)=\dot z_k(0)=0$. Let $Z^{(i)}_N\equiv \sup_{t>0} |\frac{d^i}{dt^i}(z_N(t)-z_0(t))|$. The system is called `impulse stable' if it is asymptotically stable and if for $i$ equal to 0, 1 and 2: $\limsup_{N\rightarrow \infty}\; \left|Z^{(i)}_N\right|^{1/N}\leq 1$. Otherwise the system is `impulse unstable'.
\label{defn:impulse}
\end{defn}
Impulse instability in this sense means that if we give the leader a 'unit-kick', then that perturbation travels through the flock and causes $\sup|z_N(t)|$, $\sup |\dot z_N(t)|$, or $\sup |\ddot z_N(t)|$ to grow exponentially in $N$, before eventually dying out.

\vskip .2in
\begin{conj}
The system of Equation (\ref{eq:indepleader}) is impulse stable if and only if $\rho=\frac12$ (or $\kappa=1$).
\end{conj}

The fact that for $\rho=\frac12$ the system is stable in this sense was proved in \cite{flocks5}. When $\rho>\frac12$, Corollary \ref{cor:orbit} implies that for an amount of time $O(\kappa^{\frac{N}{2}})$ the trailing agent appears to not move. During all that time however the leader has traveled at unit speed. Thus at this point $t$ is in time $z_N(t)-z_0(t)$ is exponentially large in $N$. This proves impulse instability when $\rho>\frac12$.

The problem resides in the case $\rho<\frac12$. Due to Corollary \ref{cor:many-poles} we cannot easily find approximate solutions. Here is a heuristic argument in that case. Use  the fact that $\mu_-\mu_+=\kappa$ to rewrite
\begin{equation}
a_N(\nu)=\frac{1+\kappa}{\kappa}\;\mu_-^N\;\frac{\mu_+-\mu_-}{\mu_+-\mu_+^{-1}} \;\left(1- \frac{\mu_--\mu_-^{-1}}{\mu_+-\mu_+^{-1}} \left(\frac{\mu_-}{\mu_+}\right)^N\right)^{-1} \quad
\label{eq:aN-to-mu}
\end{equation}
Now one shows (see the appendix of \cite{flocks6}) that there is an $r\in(0,1)$ for which $\frac{|\mu_-(i\omega)|}{|\mu_+(i\omega)|}<r$ and furthermore that there is an interval $(0,\omega_+)$ on which $|\mu_-(i\omega)|>1$. From this it follows that $a_N$ grows exponentially large (in $N$) as $N$ tends to infinity on a fixed interval $(0,\omega_+)$. Thus the $L_2$ norm $\parallel a_N(i\omega)\parallel_2$ of $a_N$ grows exponentially. Since the Fourier transform (and its inverse) preserve the $L_2$ norm (by Plancherel's theorem), we now have that $\ddot z_N$ has an exponentially growing $L_2$ norm. Until this point there is no problem. But now we want to prove that the sup-norm of $\ddot z_N(t)$ must also grow exponentially.
We know that $z_N$ is a linear combination of eigensolutions each of which decays with $e^{Re(\nu_{\ell\pm})t}$. In this case the numbers
$Re(\nu_{\ell\pm})$ are uniformly (in $N$) bounded by a strictly negative number. So it \emph{seems} reasonable that \emph{if}
such a function is to have an exponentially large $L_2$ norm then its sup-norm must grow exponentially as well. However, we have been unable to prove this rigorously.

\vskip .5in
\section*{Acknowledgements:} JJPV is grateful for useful conversations with Gerardo Lafferriere.

\vspace{\fill}

\end{document}